# Chain Flexibility and the Segmental Dynamics of Polymers

Daniel Fragiadakis*, C. Michael Roland

*Chemistry Division, Naval Research Laboratory, Washington, DC, 20375-5342, USA.*

**ABSTRACT**. Using molecular dynamics simulations we examine the dynamics of a family of model polymers with varying chain length and torsional potential barriers. We focus on features of the dynamics of polymers that are seen experimentally but absent in simulations of freely rotating and freely jointed chains. The reduced effect of volume on the segmental dynamics with increasing chain length, a capacity for pressure densification, and the deviation from constant Johari-Goldstein relaxation time at constant segmental relaxation time all have a common origin — torsional rigidity, and these effects become increasingly apparent for more rigid chains.

**INTRODUCTION**

With respect to their segmental dynamics, polymers in the supercooled regime are virtually indistinguishable from their small molecule counterparts. At frequencies beyond terminal flow and the Rouse modes, the segmental relaxation of polymers have spectral shapes, temperature and pressure dependences, physical aging behavior, and dynamic heterogeneity that are qualitatively equivalent to those properties of molecular liquids. The class of "simple" liquids, per the isomorph description of Dyre and coworkers[1], encompassing non-associated molecular liquids also seems to include most polymers.[2] Both polymers and molecular liquids conform to density scaling; that is, the relaxation time $\tau_\alpha$ and transport properties can be expressed as a function of the ratio of temperature and density, the latter raised to a material constant $\gamma$: $\tau_\alpha = f(\rho^\gamma/T)$. The exponent $\gamma$ is related to the steepness of the intermolecular potential and quantifies the relative influence of temperature and volume on dynamics[3,4,5,6,7].

While the flexibility of polymers chains can have a profound effect on certain physical properties, motion encompassing many repeat units is largely independent of chemical structure[8]. Although introducing bending and torsional potentials slows down local relaxation, increasing the glass transition temperature, and affects the degree of chain entanglement, the qualitative nature of the dynamics does not change[9,10]. Thus in computer simulations investigating polymer dynamics, freely jointed or freely rotating chains of Lennard-Jones particles are widely used[11]. Like their small molecule counterparts, these simple models are attractive since they are much more computationally efficient than atomistic models, but also lack chemical specificity and therefore might provide a more general understanding of glass-forming systems[12]. However, simulated freely jointed and freely rotating chains lack three dynamic properties, at first glance unrelated, that are observed experimentally for real polymers:

1) The density scaling exponent $\gamma$ decreases with chain length[13,14,15]; i.e., the dynamics in higher molecular weight polymers is less sensitive to volume.
2) Polymers pressure-densify: vitrification under pressure yields higher density than conventional glasses formed at ambient pressure.[16,17]
3) The Johari-Goldstein secondary relaxation time, $\tau_{JG}$, typically constant at fixed $\tau_\alpha$ for molecular liquids, becomes smaller for polymers at constant $\tau_\alpha$[18,19].

**COMPUTATIONAL DETAILS**

We simulate a family of chains based on Lennard-Jones particles with the size parameter and potential well depth set to unity. Bond lengths and angles are kept constant to within a few percent using harmonic potentials: neighboring segments are linked by harmonic bonds with length 0.5 and force constant $10^5$; for the bond angle potential the equilibrium angle was 120° with spring constant 1000. We determined that the results are qualitatively independent of the force constants used (values of $10^2$-$10^4$ were tested). In addition, for chain lengths N>3 a sinusoidal torsional potential was used to introduce chain rigidity:

$$U(\theta) = 0.5A(1 + \cos 3\theta) \qquad (1)$$

where $\theta$ is the dihedral angle and A the barrier height. For a freely rotating chain, $A = 0$. For a freely jointed chain, the angle potential is also absent. In this work we vary $A$ to investigate specifically chain flexibility, i.e., the effect that the torsional potential governing rotation about backbone bonds had on the supercooled dynamics, in particular the three properties enumerated above.

Approximately 8000 Lennard-Jones particles (the closest multiple of the chain length) were simulated for each system. MD simulations were carried out in the NPT ensemble, using the RUMD software[20] using a Nosé-Hoover thermostat[21]; the software was also modified to incorporate a Berendsen barostat[22]. For the heating and cooling ramps employed to study pressure densified systems, we set the barostat time constant to be short enough (determined by trial and error) that the target pressure was accurately maintained throughout.

We consider a system fully equilibrated when the chain end-to-end correlation function has decayed to near zero. For long chains and low temperatures it is sometimes impractical to wait for full equilibration. In those cases we first fully equilibrate a system at the target density and higher temperature, and then decrease the temperature and equilibrate for a time at least 100 times longer than the segmental relaxation time. We determined that that is sufficient for the static (density or pressure, potential energy) and dynamic properties ($\alpha$ and end-to-end relaxation times) to no longer exhibit any time dependence.

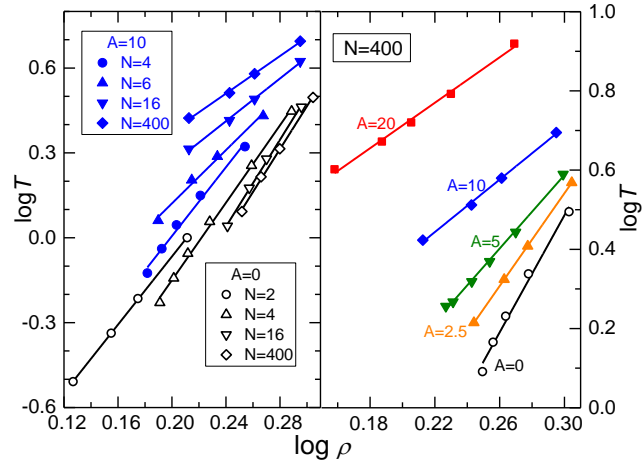

**Figure 1.** Double logarithmic plot of the temperature and density along an isochrone with $\tilde{\tau}_\alpha \simeq 10^3$ for the indicated simulated polymers. The slope of the fitted lines equals the density scaling exponent γ.

For selected state points of the N=100 and N=400 systems, we ran simulations of a larger system (32000 particles) to assess size effects, and found no significant difference in the segmental dynamics.

**RESULTS AND DISCUSSION**

Along the glass transition or any other line of constant relaxation the density scaling property can be expressed as[23]

$$T\rho^{-\gamma} = \text{constant} \qquad (2)$$

More accurately, the reduced relaxation time $\tilde{\tau}_\alpha$ [24, 25] should be used, defined as

$$\tilde{\tau}_\alpha = v^{-1/3}(kT/m)^{1/2}\tau_\alpha \qquad (3)$$

where $m$ and $v$ are the molecular mass and volume.

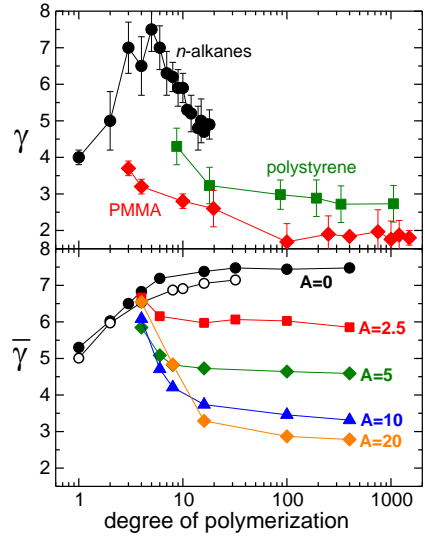

**Figure 2.** (upper) Experimental density scaling exponent for polymers as a function of the number of repeat units. *N*-alkane data is from ref. [24]; PMMA PVT data from ref. [13] and reanalysis of data from [28,29,30,31]; polystyrene from analysis of PVT data [30,32,33]. (lower) Average density scaling exponent vs. number of repeat units for the simulated polymers having the indicated torsional potential barrier *A*. The empty symbols are for the freely jointed chain of ref. [26].

Thus, double logarithmic plots of *T* versus inverse *ρ* at constant reduced relaxation time should be straight lines having a slope equal to $\gamma$. Such plots are shown in Figure 1 for chains of various length *N* and torsional barrier height *A*; the sets of state points with constant relaxation time were found simply by trial and error. The data for the most and least flexible chains with *N*=400 in Fig. 1 have some curvature; that is, $\gamma$ is weakly density dependent.

In Figure 2 are plotted the mean $\gamma$ as a function of chain length for values of *A* from 0 to 20. For *A*=0 (freely rotating chains) $\gamma$ increases systematically with increasing chain length and saturates at N=20. The values of $\gamma$ are similar to those found by Veldhorst[26] for a freely-jointed chain with rigid bonds longer by a factor of 2. The interpretation was that the rigid bonds can be seen as an infinitely steep repulsive potential that increases the effective steepness of the intermolecular potential. We recently found that the increase in $\gamma$ from atomic to molecular and polymeric liquids based on the same interatomic potential can be quantitatively related to the molecular shape, which determines the scaling of the relevant intermolecular distance with density[27].

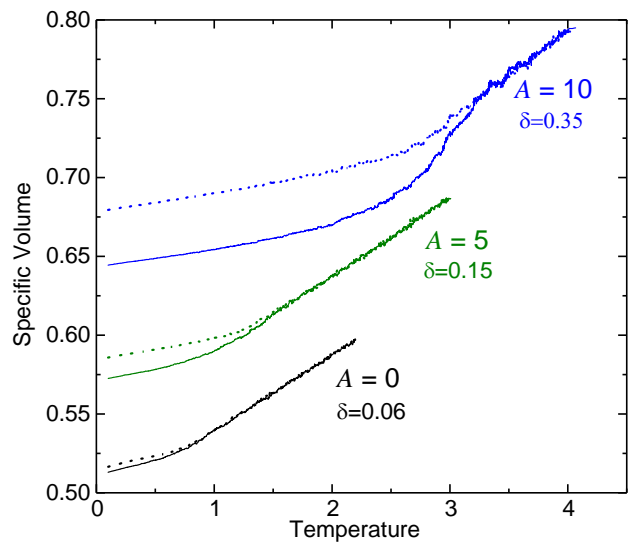

**Figure 3**. Volume at $P_0 = 1$ during heating of glass formed at $P_1=1$ (dotted lines) and $P_1=10$ (solid lines) for a freely rotating and two semiflexible polymers. Curves have been shifted vertically for clarity. Cooling and heating rates are $2.5 \times 10^{-4}$ in Lennard-Jones units.

An increasing exponent is opposite of the behavior seen experimentally in polymers, where $\gamma$ decreases with molecular weight up to an asymptotic limit[13,14,15], leading to generally lower values of $\gamma$ for polymers than for molecular liquids. Also included in Fig. 2 are collected experimental results for polystyrene, polymethylmethacrylate and *n*-alkanes. These include new values from reanalysis of previously reported volumetric data, where $\gamma$ is determined from the density $\rho_g$ and temperature $T_g$ at the glass transition dependence, which are related by $T_g \rho_g^{-\gamma} = const$. We find $\gamma$ decreases with molecular weight, the only exception being the very short *n*-alkanes, for which an *increase* in $\gamma$ is seen from $N=1$ to 5. Introducing stiffness into the simulated polymers, the dependence on *N* is decreasing, in accord with the experimental behavior. This decrease is steeper for stiffer chains, and levels off at a lower value of $\gamma$ and at a higher chain length. A torsional potential only exists for $N>3$, so a maximum in $\gamma$ occurs at N=3 or 4, depending on chain rigidity, similar to the behavior of the *n*-alkanes.

Another property common to polymers is pressure densification (PD)[16,17]. This refers to formation of a glass by pressurizing the material prior to cooling below $T_g$ and releasing the pressure; conventional glass (CG) is formed by simply cooling at ambient pressure. The PD glass has higher density and other property differences from CG. The freely jointed polymer cannot be pressure densified; its density is equivalent to that of the glass formed at low pressure[34]. In Figure 3 are shown for simulated chains having varying flexibility the heating curves for the glass formed at low and high pressure. A measure of the effect of pressure densification is the dimensionless quantity $\delta$ defined as[34]

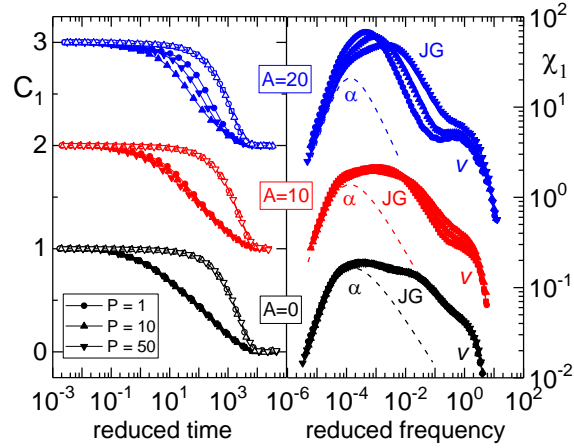

**Figure 4.** Dynamics in reduced units for the freely rotating and two semi-flexible chains with length N=16 at the indicated pressures and at temperatures for which $\tilde{\tau}_\alpha \simeq 10^3$ (along an $\alpha$ relaxation isochrone). Left: rotational correlation function $C_1$ (solid symbols) and end-to-end correlation function (empty symbols). Right: Imaginary part of the susceptibility corresponding to $C_1$. Maxima for the α and JG relaxations and vibrational modes are evident; the contribution of the α relaxation to each P=0 spectrum is shown as a dotted line. For this chain length the end-to-end relaxation does not result in a separate peak in $\chi_1$.

$$\delta = \frac{v_{CG} - v_{PD}(P_0)}{v_{CG} - v_{PD}(P_1)} \qquad (4)$$

where the specific volume $v_{CG}$ of the CG is measured at low pressure $P_0$, and that of the PD glass at $P_0$ and at the (higher) densification pressure $P_1$. For polymers $\delta$ spans values from 0.13 to 0.36.[34,35] The $\delta$ in Fig. 3 are in this range for chains with $A \geq 5$. More flexible chains show small or no effect of vitrification pressure; for a freely jointed chain (not shown) we find δ=0.

Also seen in Fig. 3 is that the glass transition temperature, defined as the temperature at which the thermal expansion coefficient begins to increase during heating. $T_g$ is always lower for the PD glass. Thus, by either of the two criteria, the extent of densification or the stability of the glass, stiffer chains exhibit a larger effect of pressure densification, behavior comparable to real polymers.

The last property we investigate is isochronal superpositioning of the relaxation dispersion for the segmental dynamics and for the Johari-Goldstein secondary process. The Johari-Goldstein (JG) relaxation is a specific type of secondary relaxation, present in a wide variety glass-forming liquids and polymers, which involves motion of all atoms in the molecule or polymer repeat unit[36] and some or all molecules depending on the material and temperature[37]. Several properties of the JG relaxation are related to the main α relaxation[38]; one of those relationships is that for many molecular liquids and their mixtures $\tau_{JG}$ is constant for constant $\tau_\alpha$.[39,40] For polymers, however, $\tau_{JG}$ decreases for constant $\tau_\alpha$ with increasing $T$ and $P$.[18,19] In Figure 4 are shown the first order bond rotational correlation function, $C_1(t) = \langle\cos\theta(t)\rangle$ (where $\theta(t)$ is the angular change of a bond vector) and its Fourier transform $\chi_1$, as well as the chain end-to-end correlation function. At each pressure the temperature is determined that gives superposition of the segmental relaxation, in order to identify state points having constant $\tau_\alpha$. It is seen that a corresponding collapse of the full spectrum, which includes the JG relaxation, is obtained only for the freely rotating chain. There is no collapse for $A>0$ due to the JG relaxation having a different ($T,P$) dependence. This breakdown of isochronal superpositioning for less flexible chains is brought out more clearly in the susceptibility spectra (right panel of Fig. 4). The spectra were fit to a sum of three contributions, the local segmental dynamics, the JG process, and the vibrations. The segmental relaxation was described using a Kohlrausch function, with $\tau_\alpha$ and the stretch exponent kept fixed. The latter follows from the invariance of the $\alpha$ relaxation function at fixed relaxation time.[41,42]

As evident in the spectra in Fig. 4, for $A>0$ there is a systematic increase in the frequency of the JG peak with increasing $T$ and $P$. This breakdown of isochronal superpositioning corresponds to the behavior seen in experiments on polymers.[17,18,19]

**CONCLUSIONS**

Molecular dynamics simulations of polymer chains reproduce the segmental dynamic properties of real polymers, provided there are constraints on the rotation about backbone bonds. The diminished effect of volume on segmental dynamics at higher chain length, the capacity for pressure densification, and deviation from constant JG relaxation time at constant $\tau_\alpha$ all arise from torsional rigidity and become increasingly apparent for more rigid chains. Simulations of more flexible polymers, including freely jointed and freely rotating chains that lack a torsional potential, exhibit results for the $\alpha$ relaxation that do not match experiments. The origin of the discordant behavior is the capacity of flexible chains to mitigate intermolecular constraints on their local motions, in this respect mimicking to some extent molecular liquids, with which they share the aforementioned properties. A similar effect can be seen in the fragility (steepness in the temperature dependence of segmental relaxation) which is broadly in the same range as molecular liquids for flexible polymers with no side groups but typically higher for other polymers[43]. For polymers with side groups, the relative stiffness of side groups and main chain may control fragility[44], and it would be interesting to examine the effect of side groups on the three properties considered herein.

**ACKNOWLEDGMENT**

This work was supported by the Office of Naval Research.